\documentclass[twocolumn,twoside,preprintnumbers,amsmath,
amssymb,showkeys]{revtex4}

\usepackage{graphicx}
\usepackage{fancyhdr}
\usepackage{pslatex}
\usepackage{amsmath}
\usepackage{amssymb}

\newcommand{\lleq}[1]{\label{#1} }
\def\pp{}
\newcommand{\inggr}[1]{\includegraphics[#1]}
\newcommand{\citt}[1]{\cite{#1}}
\def\bp {{ \mathbf{p} }} 
\def\bq {{ \mathbf{q} }} 
\def\bK {{ \mathbf{K} }} 

\begin{document}

\title{HBT shape analysis with $\bq$-cumulants}

\author{H.C.\ Eggers$^1$ and P.\ Lipa$^2$}

\affiliation{$^1$ Department of Physics,
  University of Stellenbosch, 7602 Stellenbosch, South Africa\\
  $^2$ Arizona Research Laboratories,
  Division of Neural Systems, Memory and Ageing,
  University of Arizona, Tucson AZ 85724, USA.}

\begin{abstract}
  Taking up and extending earlier suggestions, we show how two- and
  threedimensional shapes of second-order HBT correlations can be
  described in a multivariate Edgeworth expansion around gaussian
  ellipsoids, with expansion coefficients, identified as the cumulants
  of pair momentum difference $\bq$, acting as shape parameters.
  Off-diagonal terms dominate both the character and magnitude of
  shapes. Cumulants can be measured directly and so the shape analysis
  has no need for fitting.
\end{abstract}

\vskip -1.35cm

\maketitle

\thispagestyle{fancy}
\setcounter{page}{1}

\bigskip

\section{Introduction}

Early measurements of the Hanbury-Brown Twiss (HBT) effect made use of
momentum differences in one dimension, for example the four-momentum
difference $q_{\rm inv}$ \citt{Gol60a}. The huge experimental
statistics now available permits measurement of the effect in the
three-dimensional space of vector momentum differences $\bq = \bp_1 -
\bp_2$ and, in many cases, also its dependence on the average pair
momentum $\bK = (\bp_1 + \bp_2)/2$.  Increasing attention has
therefore been paid in the last decade to the second-order correlation
function in its full six-dimensional form,
\begin{eqnarray}
\lleq{inb} 
&C_2&\!\!\!\!\!(\bq,\bK) 
\;  = \;  1 \; + \; R_2(\bq,\bK) \\
&=& \frac{
\int d\bp_1\; d\bp_2\; 
\rho(\bp_1,\bp_2) \,
\delta(\bp_1 - \bp_2 - \bq)\,
\delta(\tfrac{1}{2}(\bp_1 + \bp_2) - \bK)
}
{
\int d\bp_1\; d\bp_2\; 
\rho^{\mathrm{ref}}(\bp_1,\bp_2) \,
\delta(\bp_1 - \bp_2 - \bq) \,
\delta(\tfrac{1}{2}(\bp_1 + \bp_2) - \bK)
},
\nonumber
\end{eqnarray}
with $\rho$ the density of like-sign pairs in sibling events and
$\rho^{\mathrm{ref}}$ the event-mixing reference. While $C_2$ data can
be visualised and quantified reasonably well in two dimensions
\citt{NA22-96a,Egg06a}, it is harder to quantify three- or
higher-dimensional correlations. Projections onto marginal
distributions are inadequate \citt{Lis05a}, while sets of conditional
distributions (``slices'') require many plots and miss cross-slice
features.
\pp

Under these circumstances, efforts to quantify the shape of the
multidimensional correlation function with Edgeworth expansions
\citt{Heg93a,Cso00a} or spherical harmonics \citt{Lis05a} represent
welcome progress. More ambitious programmes seek to extend connections
between gaussian source functions and the ``radius parameters'' of the
correlation function to sets of higher-order coefficients using
imaging techniques \citt{Bro97a,Bro98a,Bro05a} and cartesian harmonics
\citt{Dan05a}.
\pp

In this contribution, we extend the Edgeworth expansion solution
proposed in \citt{Heg93a,Cso00a} to a fully multivariate form,
including cross terms. Generically, the intention is to expand a
measured normalised probability density $f(\bq)$ in terms of a
reference density $f_0(\bq)$ and its derivatives,
\begin{equation} \lleq{inc} 
f(\bq) = f_0(\bq) 
\left\{\; \mbox{Edgeworth\
    expansion\ in\ } \bq \; \right\},
\end{equation}
so as to characterise $f(\bq)$ by its expansion coefficients.  While
we have previously made use of a discrete multivariate Edgeworth form
with poissonian reference $f_0$ to describe multiplicity distributions
\citt{Lip96a}, the shape analysis of $R_2$ requires the more
traditional continuous version \citt{Ken87a} with a gaussian reference
$f_0$.  For the purpose of analysing the shape of the experimental
correlation function in HBT, we hence define the measured nongaussian
probability density as
\begin{equation} \lleq{ind}
f(\bq,\bK) = \frac{R_2(\bq,\bK)}{\int d\bq\, R_2(\bq,\bK) },
\end{equation}
where $R_2(\bq)$ is itself a normalised cumulant of pair counts
\citt{Car90d}. For the reference distribution (null case), we take
the multivariate gaussian, which in its most general form is
\begin{equation} \lleq{npb}
f_0(\bq,\bK) = 
\frac{
\exp\left[ 
-\tfrac{1}{2} (q_i - \lambda_i) (\lambda^{-1})_{ij} (q_j - \lambda_j)
\right]
}
{
(2\pi)^{D/2} (\det\lambda)^{1/2}\;
},
\end{equation}
where $D$ is the dimensionality of $\bq$, Einstein summation
convention is used (here and throughout this paper), the $\lambda_i$
are the first ``$\bq$-cumulants'' of $f_0$ \citt{Wie96a,Wie96b},
\begin{equation} \lleq{npc} 
\lambda_i(\bK) = \textstyle\int d\bq\; f_0(\bq,\bK)\, q_i,
\end{equation}
$\lambda_{ij}(\bK)$ is the covariance matrix (the set of second-order
$\bq$-cumulants) in the components of $\bq$,
\begin{equation} \lleq{npd} 
\lambda_{ij}(\bK) 
= \left( \textstyle\int d\bq\; f_0\, q_i\, q_j \right)
- \left( \textstyle\int d\bq\; f_0\, q_i \right)
  \left( \textstyle\int d\bq\; f_0\, q_j \right),
\end{equation}
and $\lambda^{-1}_{ij}$ the inverse matrix. While we suppress $\bK$ in
our notation from now on, all results are valid for $\bK$-dependent
first moments $\lambda_i$ and covariance matrix elements
$\lambda_{ij}$.

\section{Reference distribution}

The vector difference is normally decomposed into components $\bq =
(q_1,q_2,q_3) = (q_o,q_s,q_l)$ in the usual (out, side, long)
coordinate system; for illustrative purposes we will also make use of
a two-dimensional vector $\bq = (q_1,q_2)$. (This is not the
two-dimensional decomposition into $(q_t,q_l)$ used in some
experimental HBT analyses \citt{NA22-96a,Egg06a}, because $q_t =
(q_o^2 + q_s^2)^{1/2}$ is always positive, while $(q_1,q_2)$ in our
two-dimensional example can be positive or negative.)  For the
two-dimensional decomposition, the covariance matrix
\begin{eqnarray}\lleq{npe}
\lambda =
\left(
\begin{array}{ccc}
\lambda_{11} & \lambda_{12}  \\
\lambda_{12} & \lambda_{22}  \\
\end{array}
\right)
\end{eqnarray}
has the inverse
\renewcommand{\arraystretch}{1.5}
\begin{eqnarray}\lleq{npf}
\lambda^{-1} &=&
\left(
\begin{array}{ccc}
\frac{1}{\chi \sigma_1^2} & \frac{-\rho}{\chi\sigma_1 \sigma_2} \\
\frac{-\rho}{\chi\sigma_1 \sigma_2} & \frac{1}{\chi \sigma_2^2} \\
\end{array}
\right),
\end{eqnarray}
where we have introduced standard deviations $\sigma_i =
\sqrt{\lambda_{ii}}$ as well as the Pearson correlation coefficient
$\rho = \sigma_{12}^2 / (\sigma_1\sigma_2) = \lambda_{12} /
\sqrt{\lambda_{11} \lambda_{22}} $ and $\chi = 1 - \rho^2$
\citt{Ken87a,Sch94a}. 
%
%
%
%
Similarly, in three dimensions the covariance matrix
\begin{eqnarray}\lleq{nph}
\lambda =
\left(
\begin{array}{ccc}
\lambda_{oo} & \lambda_{os} & \lambda_{ol} \\
\lambda_{os} & \lambda_{ss} & \lambda_{sl} \\
\lambda_{ol} & \lambda_{sl} & \lambda_{ll} \\
\end{array}
\right)
\end{eqnarray}
has the general inverse
\begin{widetext}
\renewcommand{\arraystretch}{1.7}
\begin{eqnarray} \lleq{nphf}
\lambda^{-1} &=&
\frac{1}{\det \lambda}
\left(
\begin{array}{ccc}
\lambda_{ll} \lambda_{ss} {-} \lambda_{sl}^2 
&
 \lambda_{ol} \lambda_{sl} {-} \lambda_{ll} \lambda_{os} 
& 
 \lambda_{os} \lambda_{sl} {-} \lambda_{ol} \lambda_{ss}
\\
 \lambda_{ol} \lambda_{sl} {-} \lambda_{ll} \lambda_{os} 
& %
 \lambda_{ll} \lambda_{oo} {-} \lambda_{ol}^2 
& 
 \lambda_{ol} \lambda_{os} {-} \lambda_{oo} \lambda_{sl}
\\
 \lambda_{os} \lambda_{sl} {-} \lambda_{ol} \lambda_{ss}
& 
 \lambda_{ol} \lambda_{os} {-} \lambda_{oo} \lambda_{sl} 
& 
 \lambda_{oo} \lambda_{ss} {-} \lambda_{os}^2 
\\
\end{array}
\right) \\
&=&
\frac{1}{\det \lambda}
\left(
\begin{array}{ccc}
\sigma_s^2 \sigma_l^2 \left( 1 {-} \rho_{sl}^2 \right)
&
 - \sigma_o \sigma_s \sigma_l^2 
   \left( \rho_{os} - \rho_{ol}\rho_{sl}  \right) 
& 
 - \sigma_o \sigma_s^2 \sigma_l 
   \left( \rho_{ol} - \rho_{os}\rho_{sl}  \right) 
\\
 - \sigma_o \sigma_s \sigma_l^2 
   \left( \rho_{os} - \rho_{ol}\rho_{sl}  \right) 
&
 \sigma_o^2 \sigma_l^2 \left( 1 {-} \rho_{ol}^2 \right)
& 
 - \sigma_o^2 \sigma_s \sigma_l 
   \left( \rho_{sl} - \rho_{os}\rho_{ol}  \right) 
\\
 - \sigma_o \sigma_s^2 \sigma_l 
   \left( \rho_{ol} - \rho_{os}\rho_{sl}  \right) 
& 
 - \sigma_o^2 \sigma_s \sigma_l 
   \left( \rho_{sl} - \rho_{os}\rho_{ol}  \right) 
& 
 \sigma_o^2 \sigma_s^2 \left( 1 {-} \rho_{os}^2 \right)
\\
\end{array}
\right).
\end{eqnarray}
where $\rho_{ij} = \sigma_{ij}^2 / (\sigma_i \sigma_j)$
and the determinant is given by
\begin{eqnarray} \lleq{npi}
\det \lambda 
&=& 
    \lambda_{oo} \lambda_{ss} \lambda_{ll}
-   \lambda_{oo} \lambda_{sl}^2
-   \lambda_{ss} \lambda_{ol}^2
-   \lambda_{ll} \lambda_{os}^2
+ 2 \lambda_{os} \lambda_{ol} \lambda_{sl}
\ =\  \sigma_o^2 \sigma_s^2 \sigma_l^2
\left( 1
-   \rho_{sl}^2
-   \rho_{ol}^2
-   \rho_{os}^2
+ 2 \rho_{os} \rho_{ol} \rho_{sl}
\right).
\end{eqnarray}
\end{widetext}
For azimuthally symmetric sources \citt{Cha95a}, $\rho_{os} =
\rho_{sl} = 0$, so that the inverse simplifies to
\begin{eqnarray}\lleq{npj}
\!\!\!\!\!
\lambda^{-1} 
&=& 
\!\!\!
\left(
\begin{array}{ccc}
\frac{1}{\chi\sigma_o^2} & 0 & \frac{-\rho_{ol}}{\chi\sigma_o\sigma_l} \\
0 & \frac{1}{\sigma_s^2} & 0 \\
\frac{-\rho_{ol}}{\chi\sigma_o\sigma_l} & 0 & \frac{1}{\chi\sigma_l^2}\\
\end{array}
\right) 
\equiv
\left(
\begin{array}{ccc}
2 R_{oo}^2 & 0 & 2 R_{ol}^2 \\
0 & 2 R_{ss}^2 & 0 \\
2 R_{ol}^2 & 0 & 2 R_{ll}^2 \\
\end{array}
\right)
\ \ \
\end{eqnarray}
\renewcommand{\arraystretch}{1.0}
\noindent
Identifying in the second part of Eq.~(\ref{npj}) the inverse cumulant
matrix with the usual radii $R_{ij}^2$ of the parametrisation $f_0
\sim \exp[ - \sum_{ij} R_{ij}^2 q_i q_j ]$, we note that the notation
$R_{ol}^2$ is misleading in that a positive covariance between the out
and long directions, $\rho_{ol} > 0$, results in a negative
$R_{ol}^2$.

\section{Multivariate Edgeworth expansion}

\subsection{Derivation}

In order to derive the Edgeworth expansion, we need to distinguish
between the moments and cumulants of $f_0(\bq)$ and $f(\bq)$
respectively. The cumulants of the reference $f_0$ have been fully
specified already: the order-1 and 2 cumulants are the set of
(initially free) parameters $\lambda_i$ and $\lambda_{ij}$
respectively, while all cumulants of order 3 or higher vanish
identically \citt{Ken87a} for the gaussian reference (\ref{npb}). For
the measured nongaussian $f(\bq)$, we denote the first- and
second-order moments as $\mu_i = \int d\bq\, f(\bq)\, q_i$, and
$\mu_{ij} = \int d\bq\, f(\bq)\, q_i q_j$, and in general
\begin{eqnarray}\lleq{clb}
\mu_{ijk\ldots} 
&=& \int d\bq\, f(\bq)\, q_i q_j q_k 
\ldots\;.
\end{eqnarray}
Cumulants $\kappa_{ijk\ldots}$ of $f(\bq)$ are found from these
moments by inverting the generic moment-cumulant relations
\citt{Ken87a}
\begin{eqnarray}\lleq{ing}
\mu_i &=& \kappa_i, \\
\lleq{inh}
\mu_{ij} &=& \kappa_{ij} + \kappa_i \kappa_j, \\
\mu_{ijk} &=& \kappa_{ijk} 
+ \kappa_i \kappa_{jk}
+ \kappa_j \kappa_{ki}
+ \kappa_k \kappa_{ij}
+ \kappa_k \kappa_j \kappa_k, \nonumber\\
\lleq{ini}
&=& \kappa_{ijk} 
+ \kappa_i \kappa_{jk} [3]
+ \kappa_k \kappa_j \kappa_k,
\end{eqnarray}
where we have introduced the notation $[3]$ to indicate the number of
index partitions, and therefore terms, of a given combination of
$\kappa$'s. The relations of order 4, 5 and 6, which we will need in a
moment, are
\begin{eqnarray}
\lleq{inj}
\mu_{ijkl} 
&=& \kappa_{ijkl} 
+ \kappa_i \kappa_{jkl} [4]
+ \kappa_{ij} \kappa_{kl} [3] \\
&+&
  \kappa_i \kappa_j \kappa_{kl} [6]
+ \kappa_i \kappa_j \kappa_k \kappa_l\,, 
\nonumber\\
\lleq{inja}
\mu_{ijklm} &=&
\kappa_{ijklm}
+ \kappa_i \kappa_{jklm} [5]
+ \kappa_{ij} \kappa_{klm} [10]
\\
&+& \kappa_i \kappa_j \kappa_{klm} [10]
+ \kappa_i \kappa_{jk} \kappa_{lm} [15]
\nonumber\\
&+& \kappa_i \kappa_j \kappa_k \kappa_{lm} [10]
+ \kappa_i \kappa_j \kappa_k \kappa_l \kappa_m\,,
\nonumber \\
\lleq{ink}
\mu_{ijklmn} &=&
\kappa_{ijklmn}
+ \kappa_i \kappa_{jklmn} [6]
+ \kappa_{ij} \kappa_{klmn} [15]
\\
&+& \kappa_i \kappa_j \kappa_{klmn} [15] 
+ \kappa_{ijk} \kappa_{lmn} [10],
+ \kappa_i \kappa_{jk} \kappa_{lmn} [60]
\nonumber \\ 
&+& \kappa_i \kappa_j \kappa_k \kappa_{lmn} [20]
+ \kappa_{ij} \kappa_{kl} \kappa_{mn} [15]
+ \kappa_i \kappa_j \kappa_{kl} \kappa_{mn} [45]
\nonumber\\
&+& \kappa_i \kappa_j \kappa_k \kappa_l \kappa_{mn} [15]
+ \kappa_i \kappa_j \kappa_k \kappa_l \kappa_m \kappa_n\,.
\nonumber
\end{eqnarray}
For identical particles, all moments and cumulants are fully symmetric
under index permutation.
\pp

The derivation of the Edgeworth expansion starts with the generic
Gram-Charlier series \citt{Ken87a,Cha67a,Bar88a}, which is expressed
in terms of differences between the measured and reference cumulants
\begin{eqnarray}\lleq{clf}
\eta_i &=& \kappa_i - \lambda_i, \\
\lleq{clg}
\eta_{ij} &=& \kappa_{ij} - \lambda_{ij}, \\
\lleq{clh}
\eta_{ijk} &=& \kappa_{ijk} - \lambda_{ijk} \qquad \mbox{etc.,} 
\end{eqnarray}
and moment-like entities $\zeta_{ijk\ldots}$ which are related to
the $\eta_{ijk\ldots}$ by the same moment-cumulant relations
(\ref{ing})--(\ref{ink}), i.e.
\begin{eqnarray}\lleq{cli}
\zeta_i &=& \eta_i = \kappa_i - \lambda_i, \\
\lleq{clj}
\zeta_{ij} &=& \eta_{ij} + \eta_i \eta_j
= (\kappa_{ij} - \lambda_{ij}) 
+ (\kappa_i - \lambda_i)(\kappa_j - \lambda_j),
\ \ \ 
\end{eqnarray}
and so on. The Gram-Charlier series
\begin{equation}
\frac{f(\bq)} {f_0(\bq) }
= 
1 + \zeta_i h_i(\bq) 
+ \tfrac{1}{2!} \zeta_{ij} h_{ij}(\bq)
+ \tfrac{1}{3!} \zeta_{ijk} h_{ijk}(\bq)
+ \ldots
\end{equation}
is an expansion in terms of the $\zeta$s and partial derivatives
\begin{eqnarray}\lleq{cll}
h_i(\bq) &=& -\, \frac{1}{f_0}\,
\frac{\partial f_0}{\partial q_i}, \\
\lleq{clm}
h_{ij}(\bq) 
&=& +\,\frac{1}{f_0}\,
\frac{\partial^2 f_0}{\partial q_i\, \partial q_j}, \\
\lleq{cln}
h_{ijk}(\bq) 
&=& -\,\frac{1}{f_0}\,
\frac{\partial^3 f_0}{\partial q_i\, \partial q_j\, \partial q_k}
\qquad \mbox{etc.},
\end{eqnarray}
which for gaussian $f_0$ are called hermite tensors; they will be
discussed below.  
\pp

The generic Gram-Charlier series is reduced to a simpler HBT Edgeworth
series in three steps.  First, the freedom of choice for the
parameters $\lambda_i$ and $\lambda_{ij}$ of the reference
distribution (\ref{npb}) allows us to set these to the values obtained
from the measured distribution; i.e.\ we are free to set $\lambda_i
\equiv \kappa_i$ and $\lambda_{ij} \equiv \kappa_{ij}$, so that
$\eta_i = \eta_{ij} = 0$ and hence $\zeta_i = \zeta_{ij} = 0$.
\pp

Second, we make use of the fact that all cumulants of order 3 or
higher are identically zero for the gaussian distribution,
$\lambda_{ijk} = \lambda_{ijkl} = \cdots = 0$, so that $\zeta_{ijk} =
\kappa_{ijk}$, $\zeta_{ijkl} = \kappa_{ijkl}$, $\zeta_{ijklm} =
\kappa_{ijklm}$ and in sixth order $\zeta_{ijklmn} = \kappa_{ijklmn} +
\kappa_{ijk}\kappa_{lmn}[10]$.
\pp

Finally, the contribution to the correlation function $C_2(\bq)$ of
the momentum difference $\bq_{\alpha\beta} = \bp_{\alpha{=}1} -
\bp_{\beta{=}2}$ of a given pair of identical particles
$(\alpha,\beta)$ is always balanced by an identical but opposite
contribution $\bq_{\beta\alpha} = \bp_{\beta{=}1} - \bp_{\alpha{=}2} =
- \bq_{\alpha\beta}$ by the same pair, so that $C_2(\bq)$ must be
exactly symmetric under ``$\bq$-parity'',
\begin{equation} \lleq{clo}
C_2(-\bq) = C_2(\bq).
\end{equation}
This implies that all moments and cumulants of odd order of the
measured $f(\bq)$ must be identically zero, $\kappa_{ijk} =
\kappa_{ijklm} \equiv 0$, so that terms of third and fifth order and
the $\kappa_{ijk}\kappa_{lmn}$ contribution to sixth order are also
eliminated.  
\pp

The end result of these simplifications is a multivariate Edgeworth
series in which only terms of fourth and sixth order survive,
\begin{equation}
\lleq{dgd}
\frac{f(\bq)} {f_0(\bq)}
=
1 + \tfrac{1}{4!}\, \kappa_{ijkl}\, h_{ijkl}(\bq)
+ \tfrac{1}{6!}\, \kappa_{ijklmn}\, h_{ijklmn}(\bq)
+ \ldots\,.
\end{equation}
For three-dimensional $\bq$, there are 81 terms in the fourth-order
sum and 729 in sixth order, but due to the symmetry of both the
$\kappa$ and $h$, many of these are the same. Defining $n = n_1 + n_2
+ n_3$, we introduce the ``occupation number'' notation
\begin{eqnarray}\lleq{dge}
H_{n_1 n_2 n_3} &=& 
\frac{1}{f_0}\;
\frac{(-1)^n  \;\partial^{\,n} f_0}
{
(\partial q_1)^{n_1}
(\partial q_2)^{n_2} 
(\partial q_3)^{n_3}
}, 
\end{eqnarray}
and correspondingly define cumulants $C_{n_1 n_2 n_3}$ as $\kappa_{i_1
  i_2 \cdots i_n}$ with $n_1$ occurrences of the index $1$, $n_2$
occurrences of $2$, and $n_3$ occurrences of $3$ in $(i_1 \cdots
i_n)$, e.g.\ $C_{121} = \kappa_{1223} = \kappa_{3122} = \ldots$. Similar
definitions hold for $H_{n_1 n_2}$ and $C_{n_1 n_2}$ for the
two-dimensional case.  Combining terms in (\ref{dgd}), we obtain for
the two- and three-dimensional cases respectively,
\begin{eqnarray}
\lleq{dgh}
\frac{ f(\bq) }{ f_0(\bq) }
&=&
1 + \tfrac{1}{4!}\, 
\bigl\{
    C_{40} H_{40} [2]
+ 4 C_{31} H_{31} [2]
+ 6 C_{22} H_{22}
\bigr\} 
\nonumber\\
&& \;\;\;
{}+ \tfrac{1}{6!}\, 
\bigl\{
     C_{60} H_{60} [2]
+  6 C_{51} H_{51} [2]
\nonumber\\
&& \qquad\quad
{}+ 15 C_{42} H_{42} [2]
+ 20 C_{33} H_{33}
\bigr\}
+ \ldots
\end{eqnarray}
\begin{eqnarray}
\lleq{dgg}
\frac{ f(\bq) }{ f_0(\bq) }
&=&
1 + \tfrac{1}{4!}\, 
\bigl\{
     C_{400} H_{400} [3]
+  4 C_{310} H_{310} [6]
\nonumber\\[-5pt]
&& \qquad\quad
{}+  6 C_{220} H_{220} [3]
+ 12 C_{211} H_{211} [3]
\;\bigr\}
\nonumber\\[5pt]
&& \;\;\; 
{}+ \tfrac{1}{6!}\, 
\bigl\{
      C_{600} H_{600} [3]
+   6 C_{510} H_{510} [6]
\nonumber\\
&& \qquad \quad
{}+  15 C_{420} H_{420} [6]
+  30 C_{411} H_{411} [3]
\nonumber\\
&& \qquad \quad
{}+  20 C_{330} H_{330} [3]
+  60 C_{321} H_{321} [6]
\ \ \ \ \ \ \
\nonumber\\
&&\qquad\quad {}+ 90 C_{222} H_{222}  \bigr\}
 \;\;\; + \;\;\;\ldots\,
\end{eqnarray}
where the square brackets here indicate the number of distinct
cumulants related by index permutation to those shown.  In two
dimensions, we therefore have 5 distinct cumulants of fourth order and
7 of sixth order, while in three dimensions there are 15 distinct
fourth-order and 28 sixth-order cumulants respectively. We note that
these cumulants can be nonzero even when the reference gaussian is
uncorrelated, i.e.\ even if the Pearson coefficients are zero.

\begin{figure}[!htb]
  \inggr{width=50mm}{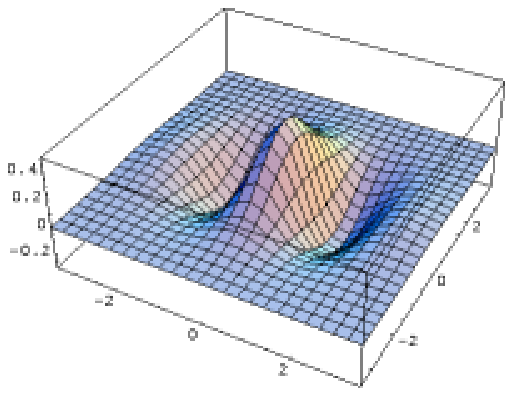}
  \\[-12pt]
  $f_0\,H_{40}$\hspace*{25mm}\mbox{}\\
  \inggr{width=50mm}{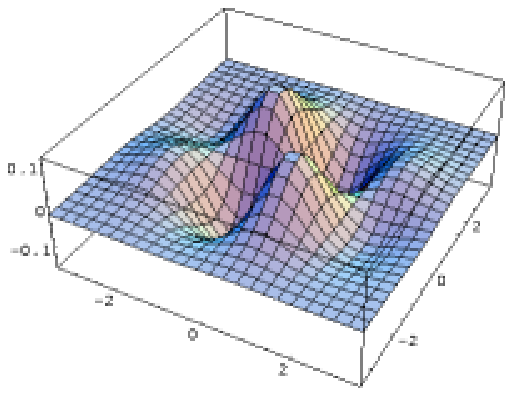}
  \\[-12pt]
  $f_0\,H_{31}$\hspace*{25mm}\mbox{}\\
  \inggr{width=50mm}{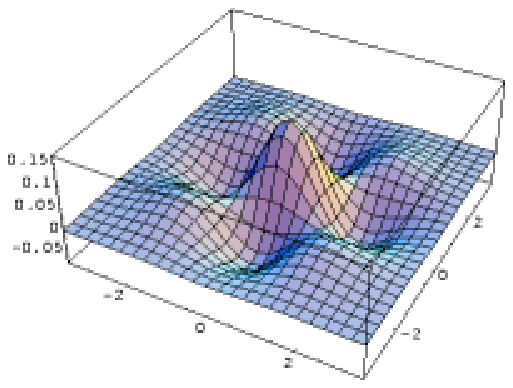}
  \\[-12pt]
  $f_0\,H_{22}$\hspace*{25mm}\mbox{}\\
  \caption{Surface plots of Gaussians times individual hermite
    tensors, $f_0\, H_{n_1 n_2}$ in two dimensions. Tensors $H_{04}$
    and $H_{13}$ are images of $H_{40}$ and $H_{31}$ mirrored through
    the $z_1 = z_2$ diagonal. Axis labels are in units of
    $\sqrt{2}\sigma_i$.}
  \ \\[-5mm]
\end{figure}

\subsection{Hermite tensors}

In the Edgeworth series (\ref{dgh}) and (\ref{dgg}), the cumulants $C$
are coefficients fixed by direct measurement, while the hermite
tensors $H$ are, through eqs.~(\ref{cll})--(\ref{cln}) and explicit
derivatives of (\ref{npb}), known functions of $\bq$.  Defining
dimensionless variables
\begin{equation} \lleq{htc}
z_i = \frac{q_i}{\sigma_i}
\end{equation}
which can also be written in terms of the usual radii as $z_i =
\sqrt{2}\, q_i\, R_{ii}$, the lowest-order hermite tensors are, for
the azimuthally symmetric out-side-long system,
\begin{eqnarray}\lleq{htd}
h_1 = H_{100} &=& \frac{z_1 - \rho z_3}{\chi \sigma_1}, \\
h_2 = H_{010} &=& \frac{z_2}{\sigma_2}, \\
h_3 = H_{001} &=& \frac{z_3 - \rho z_1}{\chi \sigma_3}.
\end{eqnarray}
Fourth-order derivatives (\ref{dge}) of the gaussian (\ref{npb})
yield, 
\begin{eqnarray}\lleq{htg}
H_{400} &=& h_1^4 - 6 h_1^2 \lambda_{11}^{-1} 
+ 3\lambda^{-1}_{11} \lambda^{-1}_{11},
\\
H_{040} &=& h_2^4 - 6 h_2^2 \lambda_{22}^{-1} 
+ 3\lambda^{-1}_{22} \lambda^{-1}_{22},
\\
H_{004} &=& h_3^4 - 6 h_3^2 \lambda_{33}^{-1} 
+ 3\lambda^{-1}_{33} \lambda^{-1}_{33},
\\
H_{310} &=& h_1^3 h_2 - 3 h_1 h_2 \lambda_{11}^{-1}, \\
H_{130} &=& h_2^3 h_1 - 3 h_1 h_2 \lambda_{22}^{-1}, \\
H_{013} &=& h_3^3 h_2 - 3 h_2 h_3 \lambda_{33}^{-1}, \\
H_{031} &=& h_2^3 h_3 - 3 h_2 h_3 \lambda_{22}^{-1}, \\
H_{301} &=& h_1^3 h_3 - 3 h_1 h_3 \lambda_{11}^{-1}
- 3 h_1^2 \lambda_{13}^{-1} + 3 \lambda_{11}^{-1} \lambda_{13}^{-1}, 
\\
H_{103} &=& h_3^3 h_1 - 3 h_1 h_3 \lambda_{33}^{-1}
- 3 h_3^2 \lambda_{13}^{-1} + 3 \lambda_{33}^{-1} \lambda_{13}^{-1}, \\
H_{220} &=& h_1^2 h_2^2 
- h_1^2 \lambda_{22}^{-1} - h_2^2 \lambda_{11}^{-1} 
+ \lambda_{11}^{-1} \lambda_{22}^{-1}, \\
H_{022} &=& h_3^2 h_2^2 
- h_3^2 \lambda_{22}^{-1} - h_2^2 \lambda_{33}^{-1}
+ \lambda_{22}^{-1} \lambda_{33}^{-1}, \\
H_{202} &=& h_1^2 h_3^2 
- h_1^2 \lambda_{33}^{-1} - h_3^2 \lambda_{11}^{-1} 
- 4 h_1 h_3 \lambda_{13}^{-1}  \nonumber\\
&&{} + \lambda_{11}^{-1} \lambda_{33}^{-1}
+ 2 \lambda_{13}^{-1} \lambda_{13}^{-1},
\end{eqnarray}
\begin{eqnarray}
H_{211} &=& h_1^2 h_2 h_3 - 2 h_1 h_2 \lambda_{13}^{-1}
- h_2 h_3 \lambda_{11}^{-1}, \\
H_{112} &=& h_1 h_2 h_3^2 - 2 h_2 h_3 \lambda_{13}^{-1}
- h_1 h_2 \lambda_{33}^{-1}, \\
H_{121} &=& h_1 h_2^2 h_3 - h_2^2 \lambda_{13}^{-1} 
- h_1 h_3 \lambda_{22}^{-1}
+ \lambda_{22}^{-1} \lambda_{13}^{-1},
\end{eqnarray}
where the inverse cumulant elements $\lambda_{ij} ^{-1}$ are functions
of the parameters $\lambda_i$ and $\lambda_{ij}$ that can be read off
from Eq.~(\ref{npj}). The differences between various permutations of
$(n_1 n_2 n_3)$ above arise from the fact that $\lambda_{12}^{-1} =
\lambda_{23}^{-1} = 0$ for azimuthal symmetry.
\pp

Note that only one of the above hermite tensors can be written in
terms of hermite polynomials at this level of generality, namely
$H_{040} = H_4(z_2)/\sigma_2^4$.  Generally, the hermite tensors
factorise into products of hermite polynomials $H_n(z_i)\ $ only if
all Pearson coefficients $\rho_{ij}$ in the gaussian reference are
zero,
\begin{equation} \lleq{hth}
H_{n_1 n_2 n_3}(\rho_{ij}{=}0) 
= \prod_{i=1}^D
\frac{ H_{n_i}(z_i) }{ \sigma_i^{n_i}}.
\end{equation}
%
\pp

\noindent
In sixth order, the tensors are generically
\begin{eqnarray} \lleq{hthb}
h_{ijklmn} &=&
h_i h_j h_k h_l h_m h_n - h_i h_j h_k h_l \lambda_{mn}^{-1}[15]
\nonumber\\
&+& h_i h_j \lambda_{kl}^{-1} \lambda_{mn}^{-1}[45]
- \lambda_{ij}^{-1} \lambda_{kl}^{-1} \lambda_{mn}^{-1}[15]
\end{eqnarray}
where again the square brackets indicate the number of distinct index
partitions. Sixth-order tensors $H_{n_1 n_2 n_3}$ can then be
constructed from these as usual, for example
\begin{eqnarray} \lleq{hthc}
H_{600} &=&
h_1^6 - 15 h_1^4 \lambda_{11}^{-1} 
+ 45 h_1^2 (\lambda_{11}^{-1})^2
- 15  (\lambda_{11}^{-1})^3
\end{eqnarray}
which closely resembles the Hermite polynomial $H_6(z) = z^6 - 15z^4 +
45 z^2 - 15$ but reduces to the latter only when $\rho_{ol} = 0$
and hence $\lambda_{11}^{-1} = 1/\sigma_1^2$.
\pp

\begin{figure}[!htb]
  \begin{minipage}[b][\height][c]{40mm}
    \inggr{width=40mm}{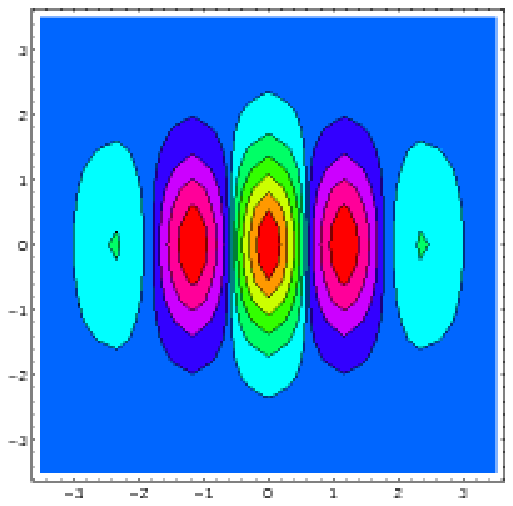}
    \centerline{$f_0\,H_{60}$}
  \end{minipage} 
  \hspace*{2mm}
  \begin{minipage}[b][\height][c]{40mm}
    \inggr{width=40mm}{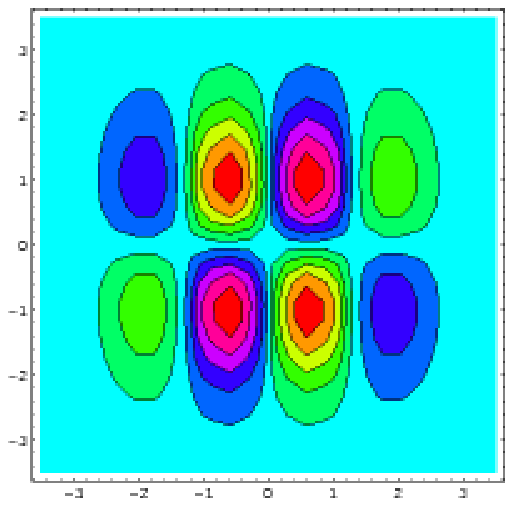}
    \centerline{$f_0\,H_{51}$}
  \end{minipage} 
  \\[8pt]
  \begin{minipage}[b][\height][c]{40mm}
    \inggr{width=40mm}{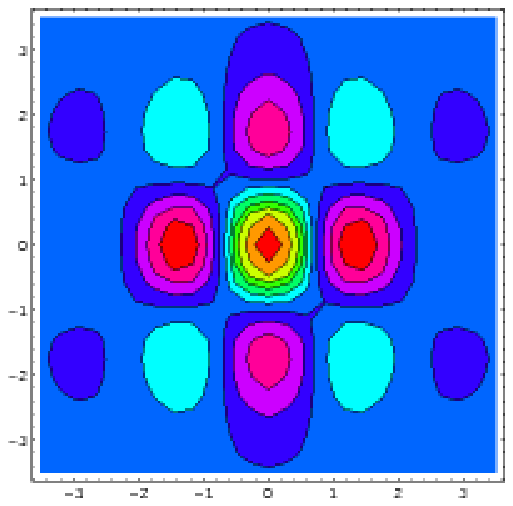}
    \centerline{$f_0\,H_{42}$}
  \end{minipage} 
  \hspace*{2mm}
  \begin{minipage}[b][\height][c]{40mm}
    \inggr{width=40mm}{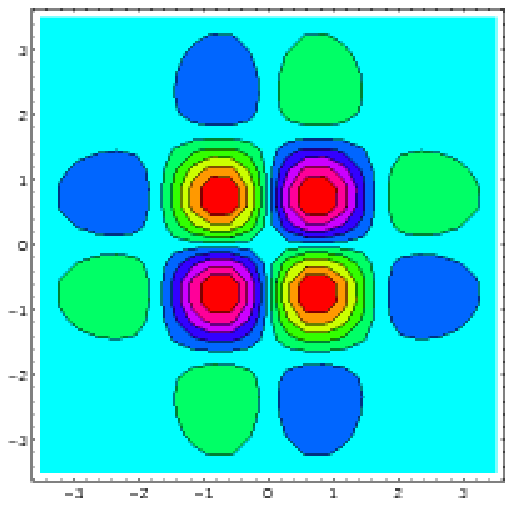}
    \centerline{$f_0\,H_{33}$}
  \end{minipage} 
  \\
  \caption{Contour plots of Gaussians times individual sixth-order
    hermite tensors, $f_0\, H_{n_1 n_2}$ in two dimensions. Red-green
    (light grey) areas represent hills while blue-red (dark grey)
    areas are valleys.  Note how regions of phase space at a distance
    of several $\sigma_i$ from the peak are probed.}
  \ \\[-5mm]
\end{figure}

\begin{figure}[!htb]
  \begin{minipage}[b][\height][c]{38mm}
    \inggr{width=38mm}{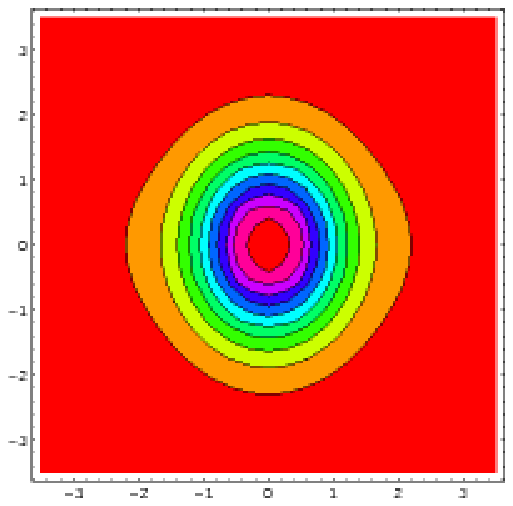}
    \centerline{$H_{40}$}
  \end{minipage} 
  \hspace*{2mm}
  \begin{minipage}[b][\height][c]{38mm}
    \inggr{width=38mm}{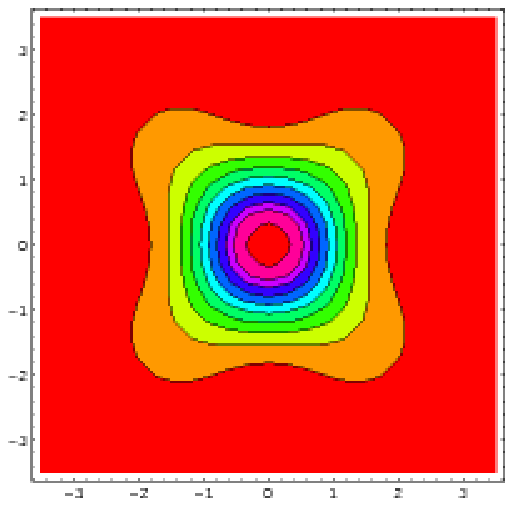}
    \centerline{$H_{22}$}
  \end{minipage} 
  \\[6pt]
  \begin{minipage}[b][\height][c]{38mm}
    \inggr{width=38mm}{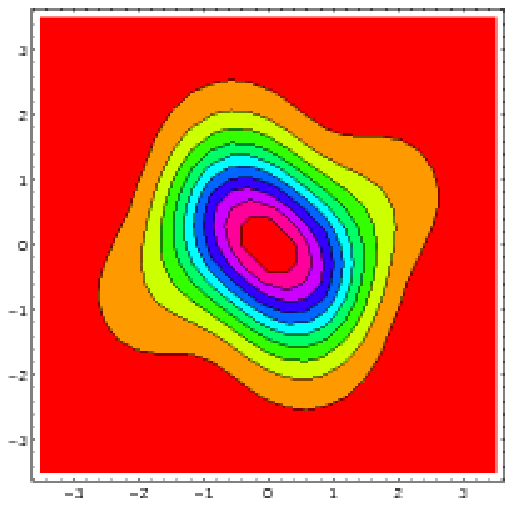}
    \centerline{$H_{31}$}
  \end{minipage} 
  \hspace*{2mm}
  \begin{minipage}[b][\height][c]{38mm}
    \inggr{width=38mm}{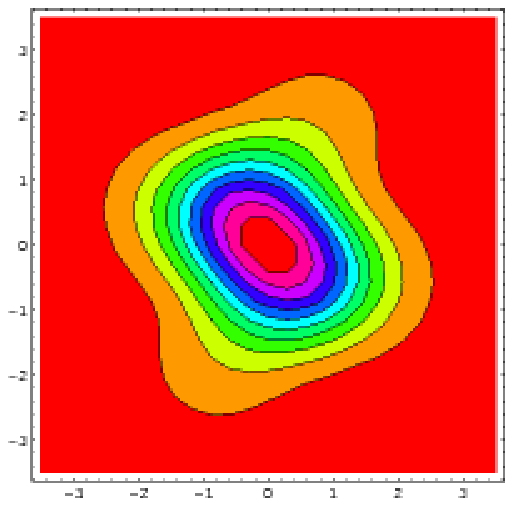}
    \centerline{$H_{13}$}
  \end{minipage} 
  \\[6pt]
  \begin{minipage}[b][\height][c]{38mm}
    \inggr{width=38mm}{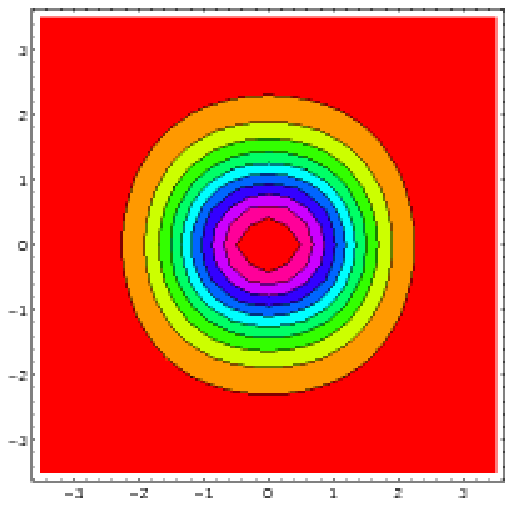}
    \centerline{$H_{60}$}
  \end{minipage} 
  \hspace*{2mm}
  \begin{minipage}[b][\height][c]{38mm}
    \inggr{width=38mm}{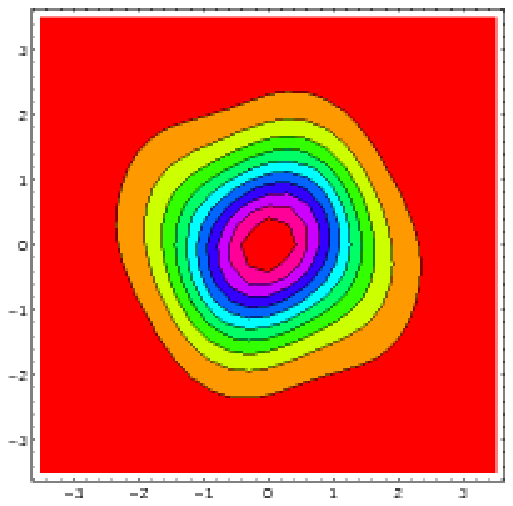}
    \centerline{$H_{51}$}
  \end{minipage} 
  \\[6pt]
  \begin{minipage}[b][\height][c]{38mm}
    \inggr{width=38mm}{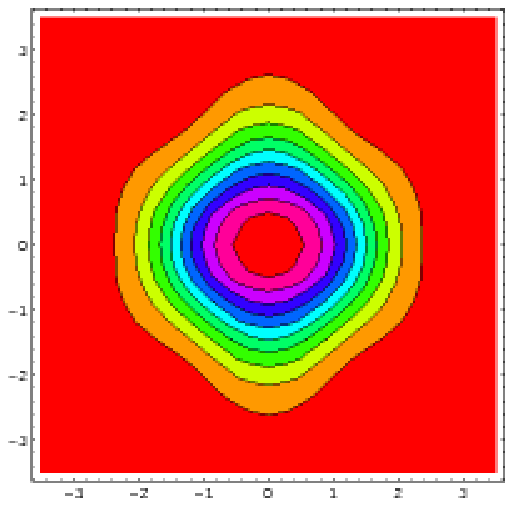}
    \centerline{$H_{42}$}
  \end{minipage} 
  \hspace*{2mm}
  \begin{minipage}[b][\height][c]{38mm}
    \inggr{width=38mm}{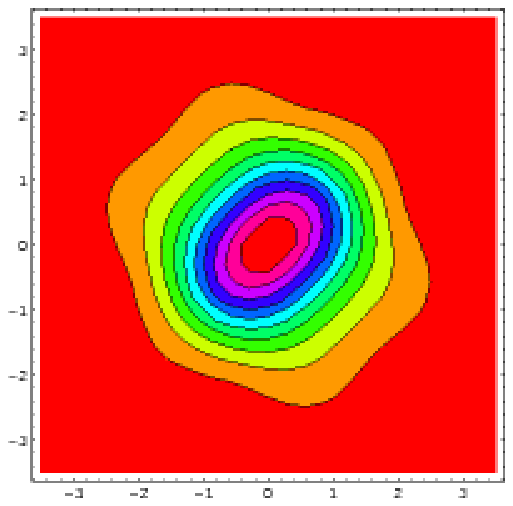}
    \centerline{$H_{33}$}
  \end{minipage} 
  \\[6pt]
  \caption{Contour plots of some individual terms in the Edgeworth
    expansion (\ref{dgh}) of the form $f_0 ( 1 + F_{n_1 n_2} C_{n_1
      n_2} H_{n_1 n_2})$ Arbitrary and unrealistically large values of
    $C_{n_1 n_2} = 1$ for fourth order and $C_{n_1 n_2} = 2$ for sixth
    order were chosen, while the combinatoric prefactors $F_{n_1 n_2}$
    are fixed in (\ref{dgh}). Both $H_{31}$ and $H_{13}$ are shown to
    exhibit their symmetry about the $z_1=z_2$ diagonal.}
  \ \\[-5mm]
\end{figure}

\begin{figure}[htb]
  \begin{minipage}[b][\height][c]{38mm}
    \inggr{width=38mm}{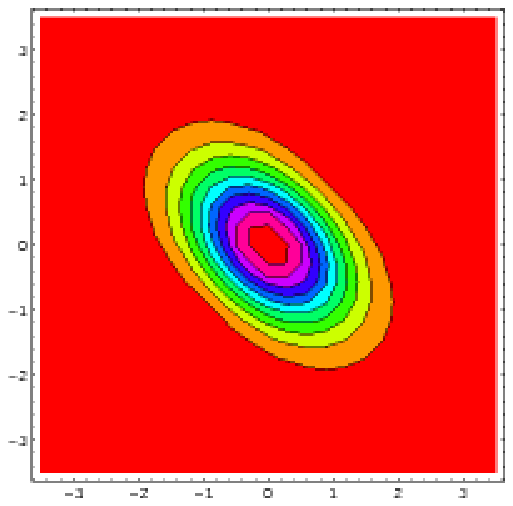}
    \centerline{(a)}
  \end{minipage}
  \hspace*{2mm}
  \begin{minipage}[b][\height][c]{38mm}
    \inggr{width=38mm}{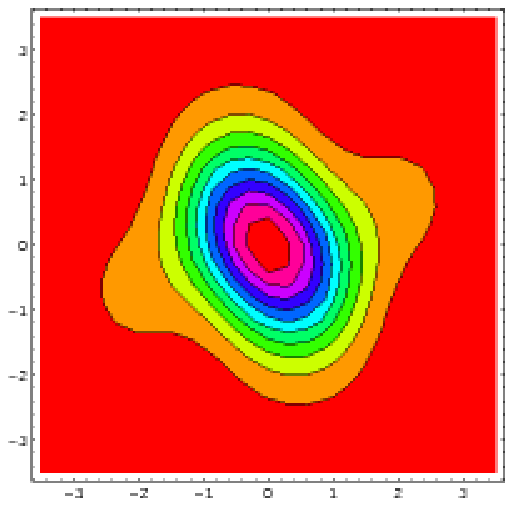}
    \centerline{(d)}
  \end{minipage}
  \\[6pt]
  \begin{minipage}[b][\height][c]{38mm}
    \inggr{width=38mm}{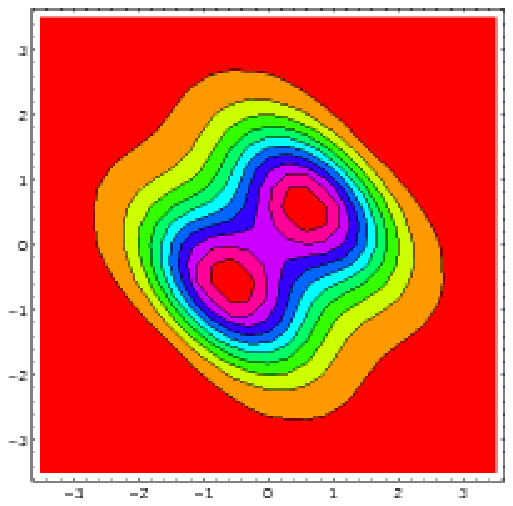}
    \centerline{(b)}
  \end{minipage}
  \hspace*{2mm}
  \begin{minipage}[b][\height][c]{38mm}
    \inggr{width=38mm}{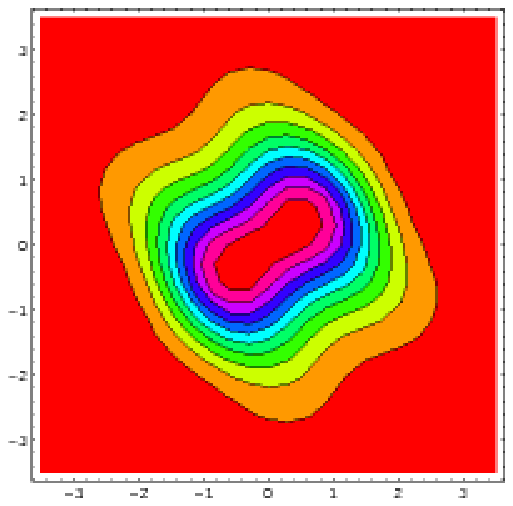}
    \centerline{(e)}
  \end{minipage}
  \\[6pt]
  \begin{minipage}[b][\height][c]{38mm}
    \inggr{width=38mm}{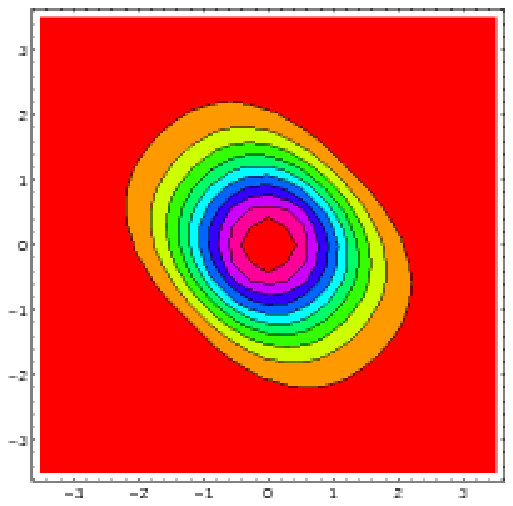}
    \centerline{(c)}
  \end{minipage}
  \hspace*{2mm}
  \begin{minipage}[b][\height][c]{38mm}
    \inggr{width=38mm}{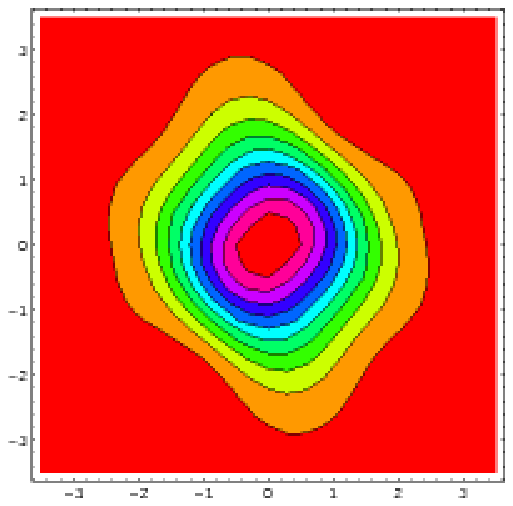}
    \centerline{(f)}
  \end{minipage}
  \\[6pt]
  \caption{Combined terms for the two-dimensional Edgeworth:
    fourth-order terms only in (a) and (d); sixth-order terms only in
    (b) and (e); and all terms in (c) and (f). Fourth-order cumulants
    were set arbitrarily to 1 and sixth-order ones to 2.  In (a)--(c),
    all cumulants are equal and nonzero, while in (d)--(f), cumulants
    $C_{04}$, $C_{13}$, $C_{06}$, $C_{15}$ and $C_{24}$ were set to
    zero in order to illustrate the possibility of asymmetric shapes.}
\end{figure}

\subsection{A gallery of shapes}

In Fig.~1, we show surface plots for individual fourth-order hermite
tensors times the two-dimensional reference gaussian, $f_0 H_{n_1
  n_2}$, with $\rho$ set to zero. As these are plotted in terms of
$\sqrt{2} z_i = \sqrt{2} q_i / \sigma_i$ \footnote{In our figures and
  Eq.~(\ref{hth}), we use for the Hermite polynomials the definition
  $H_n(x) = e^{x^2/2} (d/dx)^n e^{-x^2/2}$ which is related to the
  alternative definition $H'_n(x) = e^{x^2} (d/dx)^n e^{-x^2}$ by
  $H_n(x) = 2^{-n/2}\,H'_n(x/\sqrt{2}) $. The extra $\sqrt{2}$ factors
  creep in because \texttt{mathematica} uses the latter definition.},
the axes are scaled by the standard deviations, meaning that all
gaussians with $\rho=0$ will be circular in $(z_1,z_2)$ plots. The
individual hermite tensors clearly reflect the symmetry of their
respective occupation number indices $n_i$ and probe different parts
of the $(z_1,z_2)$ phase space as shown.  Comparing the fourth-order
terms of Fig.~1 with the sixth-order ones of Fig.~2, we note that the
latter probe regions up to several $\sqrt{2}\,\sigma_i$.  
\pp

In order to exhibit the influence of combinatoric factors, we show in
Fig.~3 individual terms of the two-dimensional Edgeworth series
(\ref{dgh}) in the form $f_0(\bq) (1 + F_{n_1 n_2} C_{n_1 n_2} H_{n_1
  n_2}(\bq))$, where the combinatoric factors $F_{n_1 n_2}$ are fixed
in (\ref{dgh}). All fourth- and sixth-order cumulants $C_{n_1 n_2}$
have been set to 1.0 and 2.0 respectively.  (This is obviously for
illustrative purposes only; in real data, smaller values are expected
and shapes will be more gaussian than those shown here.) The plots for
$H_{31}$ and $H_{13}$ illustrate the correspondence between index
permutation and symmetry about the $z_1 = z_2$ axis. We note that the
diagonal terms $H_{n0}$ have little influence on the overall shape,
while the off-diagional ones have a larger effect, not least
because of the combinatoric prefactors.
\pp

\begin{figure}[htb]
  \begin{minipage}[b][\height][c]{35mm}
    \inggr{width=35mm} {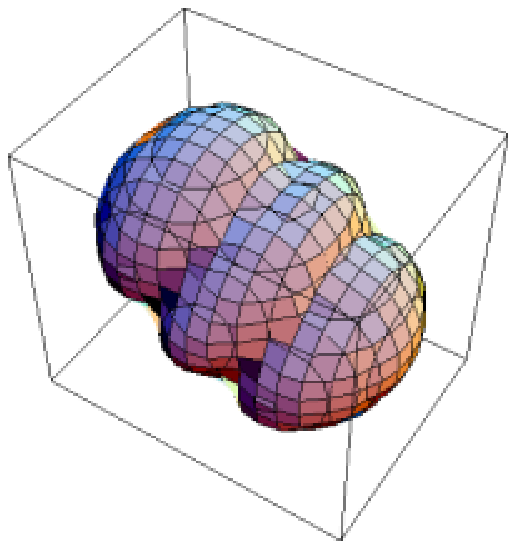}
    \\[-12pt]
    $400$\hspace*{20mm}\mbox{}\\
  \end{minipage}
  \hspace*{2mm}
  \begin{minipage}[b][\height][c]{35mm}
    \inggr{width=35mm} {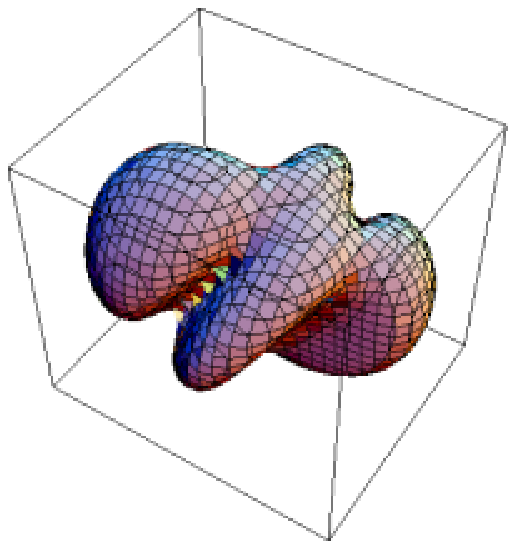}
    \\[-12pt]
    $310$\hspace*{20mm}\mbox{}\\
  \end{minipage}
  \\[6pt]
  \begin{minipage}[b][\height][c]{35mm}
    \inggr{width=35mm} {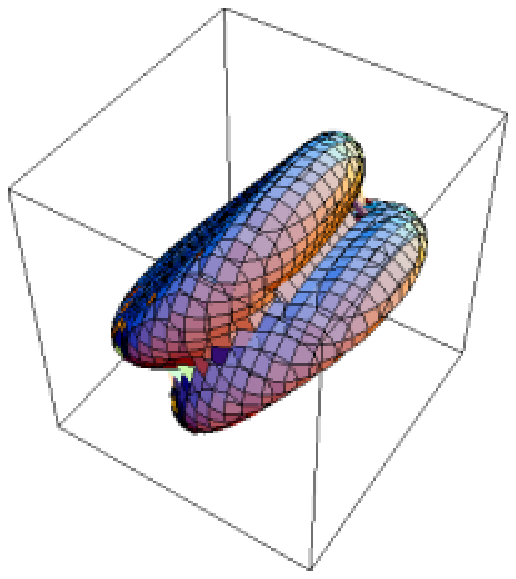}
    \\[-12pt]
    $301$\hspace*{20mm}\mbox{}\\
  \end{minipage}
  \hspace*{2mm}
  \begin{minipage}[b][\height][c]{35mm}
    \inggr{width=35mm} {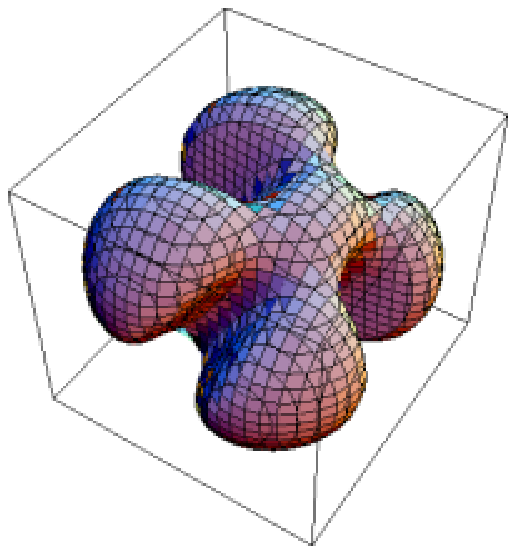}
    \\[-12pt]
    $220$\hspace*{20mm}\mbox{}\\
  \end{minipage}
  \\[6pt]
  \begin{minipage}[b][\height][c]{35mm}
    \inggr{width=35mm} {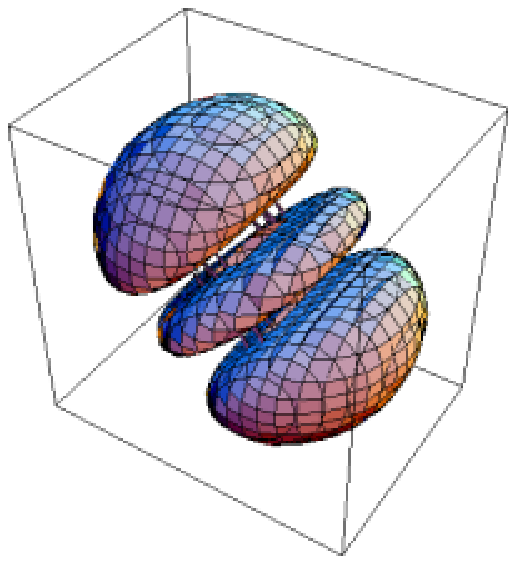}
    \\[-12pt]
    $202$\hspace*{20mm}\mbox{}\\
  \end{minipage}
  \hspace*{2mm}
  \begin{minipage}[b][\height][c]{35mm}
    \inggr{width=35mm} {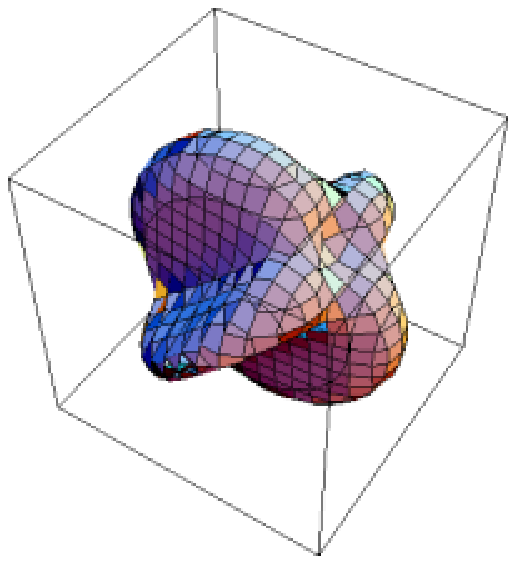}
    \\[-12pt]
    $121$\hspace*{20mm}\mbox{}\\
  \end{minipage}
  \\[6pt]
  \caption{Typical contours for three-dimensional terms $f_0(1+ F_{n_1
      n_2 n_3}\,C_{n_1 n_2 n_3}\, H_{n_1 n_2 n_3})$ with indices as
    shown.  Note that only a single contour is shown in each case.}
\end{figure}

Testing the influence of fourth- versus sixth-order terms, we show in
Fig.~4 some ``partial'' two-dimensional Edgeworth series including
only fourth-order terms, only sixth-order terms, and both orders;
again, cumulants were set to the arbitrary values of 1 and 2
respectively.
\pp

\begin{figure}[htb]
  \begin{minipage}[b][\height][c]{35mm}
    \inggr{width=35mm} {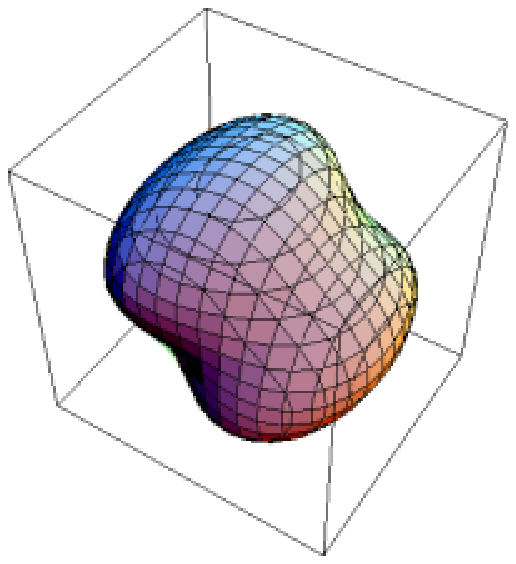}
    \\[-12pt]
    (a)\hspace*{20mm}\mbox{}\\
  \end{minipage}
  \hspace*{2mm}
  \begin{minipage}[b][\height][c]{35mm}
    \inggr{width=35mm} {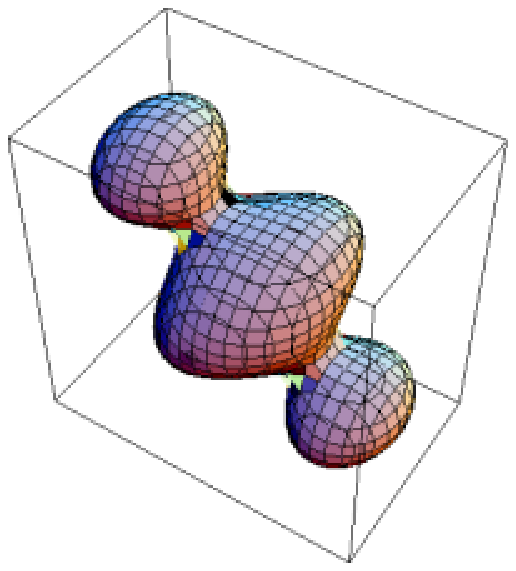}
    \\[-12pt]
    (b)\hspace*{20mm}\mbox{}\\
  \end{minipage}
  \caption{(a) Typical contour plot for combined Edgeworth with
    cumulants $C_{310}$, $C_{220}$, $C_{211}$ and their permutations
    set to 0.3 with all others (including 6th order) set to zero. (b)
    Combined Edgeworth with cumulants $C_{220}$, $C_{202}$, $C_{002}$,
    $C_{211}$, $C_{121}$, $C_{112}$ set to $0.15$ and all others to
    zero. Only single contours are shown.}
\end{figure}

In Fig.~5, a selection of shapes for individual terms $f_0(\bq)(1 +
F_{n_1 n_2 n_3} C_{n_1 n_2 n_3} H_{n_1 n_2 n_3})$ of the
three-dimensional Edgeworth expansion (\ref{dgg}) is shown. While in
the two-dimensional case full contour plots could be shown, the
surfaces shown here in each case represent only a single contour. In
Fig.~6, we show two examples with two selections of fourth-order
cumulants nonzero; the shape obviously depends strongly on their
selection and magnitude.  Clearly, effects of the different cumulants
on the overall shape often cancel out. We emphasize again that the
shapes shown are for illustrative purposes only and do not represent
real data.

\section{Discussion}

The multivariate Edgeworth expansions (\ref{dgh})--(\ref{dgg}) appear
to be a promising tool for quantitative shape analysis in HBT. While
the real test will be to gauge their performance in actual data
analysis, they do seem to have the right features and behaviour. A
number of issues deserve further comment:
\begin{enumerate}

\item It has been noted previously \citt{Wie96a,Wie96b} that the
  traditional radii of a gaussian-shaped $R_2(\bq)$ could be found
  by direct measurement rather than from fits. In the present
  formulation, this amounts to the direct measurement of the
  second-order cumulants $\lambda_{ij}$, which can be directly
  converted to ``radius parameter'' form via Eq.~(\ref{npj}).
  Going beyond Refs.~\citt{Wie96a,Wie96b}, we suggest that 
  higher-order cumulants can be measured directly also.

\item Many people have rightly expressed concern that these radii do
  not adequately represent the true shapes and behaviour of HBT
  correlations. Our Edgeworth expansion confirms that such radii are
  clearly not the whole story, but that they do represent the
  appropriate lowest-order approximation (for gaussian reference) with
  respect to which nongaussian shapes should be measured.

\item We have demonstrated that it is imperative to write Edgeworth
  expansions in a fully multivariate way: the combinatoric prefactors
  $F_{n_1 n_2 n_3}$ in (\ref{dgg}) are large for multivariate
  ``off-diagonal'' cumulants, while the influence of diagonal
  cumulants is strongly suppressed due to their small prefactors. The
  cumulant $C_{211}$, for example, has a weight 12 times larger than
  $C_{400}$, and indeed the entire expansion is dominated by the
  off-diagonal cumulants. Furthermore, even large diagonal cumulants
  do not change the shape much, as a glance at Fig.~3 will confirm.

\item Deviations from gaussian shapes are consistently quantified by
  the sign and magnitude of higher-order cumulants, which are
  identically zero for a null-case pure gaussian $f(\bq)$.  The
  Edgeworth expansion using these cumulants, while recreating the
  shape of $f(\bq)$, therefore at the same time provides a
  quantitative framework for comparison of different shapes.

\item Operationally, we suggest a procedure of successive
  approximation, whereby in a first step all elements of the
  covariance matrix $\kappa_{ij} = \lambda_{ij}$ are measured, thereby
  determining all the $\sigma$'s and the Pearson coefficient; this is
  equivalent to the usual determination of radii. This is followed by
  measurement of the set of fourth-order cumulants $C_{n_1 n_2 n_3}$.
  The measured numbers for fourth-order cumulants then represent the
  basis for shape quantification and comparison. If statistics permit,
  sixth-order cumulants can be added as a further refinement.

\item The $\bq$-cumulants proposed here are numbers rather than
  functions of $\bq$. From the viewpoint of compactness of
  description, this will be an advantage compared to the shape
  decompositions in terms of spherical and cartesian harmonics
  \citt{Lis05a,Dan05a}, in which each coefficient is a function of
  $|\bq|$. It may, however, in some cases be better to see the detail
  provided by such functions.

\item The procedure outlined above involves no fits. This represents a
  major advantage over fit-based quantification in two ways:

  Firstly: In three-dimensional analysis, typical fits are dominated
  by phase space, i.e.\ by the fact that there are many more bins at
  intermediate and large $|\bq|$ than at small $|\bq|$. This dominance
  suppresses the influence of the most interesting region on the
  $\chi^2$ for best-fit values of the parameters. In
  Ref.~\citt{Egg06a}, for example, we found that the regions of
  intermediate $|\bq|$ dominated the shape and quality of various
  fits.

  Secondly, as shown in Fig.~2, cumulants are sensitive to the tails
  of distributions, and they will hence access the same information as
  these fits and parametrisations, but in a more direct and sensitive
  way. It is well known that a direct fit to a probability
  distribution that is close to gaussian is an ineffective and
  inaccurate way to quantify nongaussian deviations, while cumulants
  do so in the most direct way possible.

\item It remains to be seen how the proposed procedure fares when the
  practical experimental difficulties of finding $f(\bq)$ and the
  higher-order cumulants come into play. Much will also depend on the
  size and accuracy of statistical errors. Fortunately, current sample
  sizes are large enough to warrant some optimism in this respect.

\item The traditional chaoticity parameter $\lambda$ remains
  undetermined within the present Edgeworth framework, because it
  cancels already in the definition (\ref{ind}) of $f(\bq)$. For a
  given level of approximation (gaussian only, fourth-order cumulants,
  sixth-order), it and the overall normalisation factor $\gamma$ may
  be recovered afterwards by using (\ref{dgg}) in a two-parameter fit
  mode using parametrisation
  \begin{equation*} \lleq{cht}
  \mbox{ } \ \ \ 
  C_2(\bq) = \gamma \left[ 1 + \lambda f_0(\bq)
             \left(\mbox{Edgeworth\ expansion}\, \right)
             \,\right]
  \end{equation*}
  with the previously experimentally-determined radii and $C_{n_1 n_2
    n_3}$ treated as constants, with $\gamma$ and $\lambda$ the fit
  parameters.

\item We note the importance of the parity argument $C_2(-\bq) =
  C_2(\bq)$ in eliminating odd-order terms in the Edgeworth expansion.
  The parity argument falls away, however, in variables where this
  symmetry does not arise; for example, any one-dimensional Edgeworth
  expansion involving only positive differences (e.g. in $q_{\rm
    inv}$) would have to include third- and fifth-order terms.

  A corollary of the parity argument is that three-dimensional
  correlations may not be represented in terms of positive absolute
  values of the components $(q_o,q_s,q_l)$ as this destroys the
  underlying symmetries. The best one can do to improve statistics is
  to combine bins that map onto each other under the transformation
  $\bq \to - \bq$ and thereby eliminate four of the eight octants in
  the three-dimensional $(q_o,q_s,q_l)$-space.

\item In the present formulation, any dependence on average pair
  momentum $\bK$ resides in the cumulants: all $\kappa_{ijk\ldots}$,
  including the traditional radii and the Pearson coefficient, are in
  principle functions of $\bK$.

\item The Edgeworth analysis set out in this contribution is based on
  a gaussian reference $f_0$. Shapes that differ significantly from
  gaussian will not be described well in either the Edgeworth
  framework or the spherical or cartesian harmonics frameworks.
  One should not, for example, expect power laws such as a pure
  Coulomb wavefunction (whose square tails off like $|\bq|^{-2}$) to
  work in a gaussian-based Edgeworth expansion.
  Indeed, it is known that large cumulants can lead to a situation
  where the truncated Edgeworth expansion of $f(\bq)$ becomes negative
  in some regions. It is therefore suitable only for shapes that do
  not deviate strongly from gaussians; for strong deviations, other
  expansions will become a necessity.

\item The Edgeworth framework is easily extended to the case of
  nonidentical particles. In that case, cumulants of all orders will
  have to be measured. It may well be that the fluctuations of
  lower-order quantities render the measurement of higher-order
  cumulants impossible, and great care will clearly have to be
  taken.

\end{enumerate}


\section*{Acknowledgments}

\noindent
This work was funded in part by the South African National Research
Foundation. HCE thanks the Tiger Eye Institute for hospitality
and inspiration.


\end{document}